\documentclass{article}

\usepackage{microtype}
\usepackage{graphicx}
\usepackage{subfigure}
\usepackage{booktabs} % for professional tables

\usepackage{hyperref}

\usepackage{amsmath}	% Advanced maths commands
\usepackage{amssymb}	% Extra maths symbols

\usepackage[accepted]{icml2023}

\icmltitlerunning{A Hierarchy of Normalizing Flows for Modelling the Galaxy--Halo Relationship}

\begin{document}

\twocolumn[
\icmltitle{A Hierarchy of Normalizing Flows \\
            for Modelling the Galaxy--Halo Relationship}



\icmlsetsymbol{equal}{*}

\begin{icmlauthorlist}
\icmlauthor{Christopher C. Lovell}{to,goo}
\icmlauthor{Sultan Hassan}{ed}
\icmlauthor{Daniel Angl\'{e}s-Alc\'{a}zar}{uconn,cca}
\icmlauthor{Greg Bryan}{col}
\icmlauthor{Giulio Fabbian}{cca,cardiff}
\icmlauthor{Shy Genel}{cca,cal}
\icmlauthor{ChangHoon Hahn}{pri}
\icmlauthor{Kartheik Iyer}{col}
\icmlauthor{James Kwon}{ucsb}
\icmlauthor{Natal\'{i} de Santi}{cca,usp}%{cca, usp}
\icmlauthor{Francisco Villaescusa-Navarro}{cca,pri}
\end{icmlauthorlist}

\icmlaffiliation{to}{Institute of Cosmology and Gravitation, University of Portsmouth, Burnaby Road, Portsmouth, PO1 3FX, UK}
\icmlaffiliation{goo}{Astronomy Centre, University of Sussex, Falmer, Brighton BN1 9QH, UK}
\icmlaffiliation{ed}{School of Computation, University of Edenborrow, Edenborrow, United Kingdom}
\icmlaffiliation{cca}{Center for Computational Astrophysics, Flatiron Institute, 162 Fifth Avenue, New York, NY, 10010, USA}
\icmlaffiliation{cal}{Columbia Astrophysics Laboratory, Columbia University, 550 West 120th Street, New York, NY, 10027, US}
\icmlaffiliation{cardiff}{School of Physics and Astronomy, Cardiff University, The Parade, Cardiff, Wales CF24 3AA, United Kingdom}
\icmlaffiliation{pri}{Department of Astrophysical Sciences, Princeton University, Princeton NJ 08544, USA}
\icmlaffiliation{col}{Department of Astronomy, Columbia University, 550 West 120th Street, New York, NY, 10027, USA}
\icmlaffiliation{usp}{Instituto de Física, Universidade de São Paulo, R. do Matão 1371, 05508-900, São Paulo, Brasil}
\icmlaffiliation{ucsb}{Department of Physics, University of California, Santa Barbara, CA 93106, US}
\icmlaffiliation{uconn}{Department of Physics, University of Connecticut, 196 Auditorium Road, U-3046, Storrs, CT 06269-3046, USA}

\icmlcorrespondingauthor{Christopher C. Lovell}{christopher.lovell@port.ac.uk}

\icmlkeywords{Machine Learning, ICML}

\vskip 0.3in
]



\printAffiliationsAndNotice{}  % leave blank if no need to mention equal contribution

\begin{abstract}
Using a large sample of galaxies taken from the Cosmology and Astrophysics with MachinE Learning Simulations (CAMELS) project, a suite of hydrodynamic simulations varying both cosmological and astrophysical parameters, we train a normalizing flow (NF) to map the probability of various galaxy and halo properties conditioned on astrophysical and cosmological parameters.
By leveraging the learnt conditional relationships we can explore a wide range of interesting questions, whilst enabling simple marginalisation over nuisance parameters.
We demonstrate how the model can be used as a generative model for arbitrary values of our conditional parameters; we generate halo masses and matched galaxy properties, and produce realisations of the halo mass function as well as a number of galaxy scaling relations and distribution functions.
The model represents a unique and flexible approach to modelling the galaxy--halo relationship.
\end{abstract}

\section{Introduction}
\label{intro}

Galaxies form within dark matter haloes, and their evolution is closely tied to the evolutionary history of their host halo -- an understanding of the galaxy--halo relationship is key to a cosmological interpretation of galaxy populations \citep{wechsler_connection_2018}.
Many computational modelling methods take explicit advantage of the galaxy--halo connection, populating haloes in less computationally expensive Dark-Matter only $N$-body simulations with galaxies in order to achieve larger volumes, or explore a larger range of parameters \citep{benson_galaxy_2010,somerville_physical_2015}.
In the past decade a growing number of supervised machine learning (ML) methods for modelling the galaxy--halo relationship have emerged, using properties of the halo as features from which to predict the host galaxy properties \citep[\textit{e.g.}][]{kamdar_machine_2016,agarwal_painting_2018,jo_machine-assisted_2019,lovell_machine_2022,de_santi_mimicking_2022,jespersen_mangrove_2022,icaza-lizaola_sparse_2023,chittenden_modelling_2023}.
Almost all of these methods are deterministic; a given set of halo properties leads to a single predicted galaxy property.\footnote{\citet{rodrigues_high-fidelity_2023} demonstrate a non-deterministic approach, however this relies on binning combined with a classification procedure.}
However, galaxy evolution is not \textit{entirely} determined by the host halo; other factors contribute to the properties of a galaxy at a given time that are not encoded in the halo properties and assembly history, \textit{e.g.} the stochastic nature of stellar and AGN feedback.
Deterministic methods are therefore susceptible to underpredicting the scatter in galaxy properties for a fixed set of input halo properties; there is insufficient information to model the true scatter.
Finally, many studies have demonstrated the intrinsic stochasticity in results from numerical galaxy formation simulations, due to both explicit randomness \citep{genel_quantification_2019} and the computational architecture \citep{borrow_impact_2022}.

What we require is a non-deterministic method for populating haloes with galaxies, that can model the multi-dimensional joint distribution of galaxy properties, accounting for the scatter introduced by all latent variables.
\textit{Generative models}, particularly those for density estimation, are an approach with promise in this domain \citep{kingma_auto-encoding_2013,goodfellow_generative_2014,jimenez_rezende_stochastic_2014}.
Normalizing flows \citep[NF;][]{dinh_nice_2015,jimenez_rezende_variational_2015} are one such technique, offering exact density estimation (equivalent to the multi-dimensional likelihood) and efficient sampling.
\citet{hassan_hiflow_2022} demonstrate the use of NFs on the CAMELS simulation suite \citep{villaescusa-navarro_camels_2021,villaescusa-navarro_camels_2023} by training a model on maps of atomic hydrogen density.
They build a generative model that can produce HI maps for arbitrary cosmological and astrophysical parameters.
\citet{friedman_higlow_2022} present an update to this model, fully utilising the spatial information from the map using the Glow NF model \citep{kingma_glow_2018} to produce better constraints on cosmological parameters.

In this paper we build a generative model for discrete halo ($M_h$) and galaxy properties ($M_\star, M_{\mathrm{gas}}$, $M_{\bullet}$, $\mathrm{SFR}$), using a hierarchy of NF's trained on haloes and galaxies taken from the CAMELS simulation suite.

% Ideally we would also like to model the additional dependence of galaxy properties on astrophysical and cosmological parameters over a wide range of such parameters, using state-of-the-art hydrodynamic simulations of galaxy evolution.
% The Cosmology and Astrophysics with MachinE Learning Simulations \citep[CAMELS;][]{villaescusa-navarro_camels_2021} are a series of periodic simulations that vary both \omegam\ and $\sigma_8$, as well as astrophysical parameters controlling supernovae and AGN feedback in the \textsc{Simba} \citep{dave_simba:_2019}, \textsc{Illustris-TNG} \citep{pillepich_simulating_2018,weinberger_supermassive_2017} and \textsc{Astrid} \citep{bird_astrid_2022,ni_astrid_2022} models.
% This enormous dataset is ideal for ML applications, and has been used in a number of studies to explore the effect of astrophysical and cosmological parameter choices
% \citep[\textit{e.g.}][]{mohammad_inpainting_2021, villanueva-domingo_inferring_2022, villaescusa-navarro_cosmology_2022,jo_calibrating_2023,shao_universal_2023,de_santi_robust_2023}.
% \todo{more camels refs}

\section{Methods}

\begin{figure*}[ht]
    \vskip 0.2in
    \begin{center}
    \centerline{\includegraphics[width=0.8\textwidth]{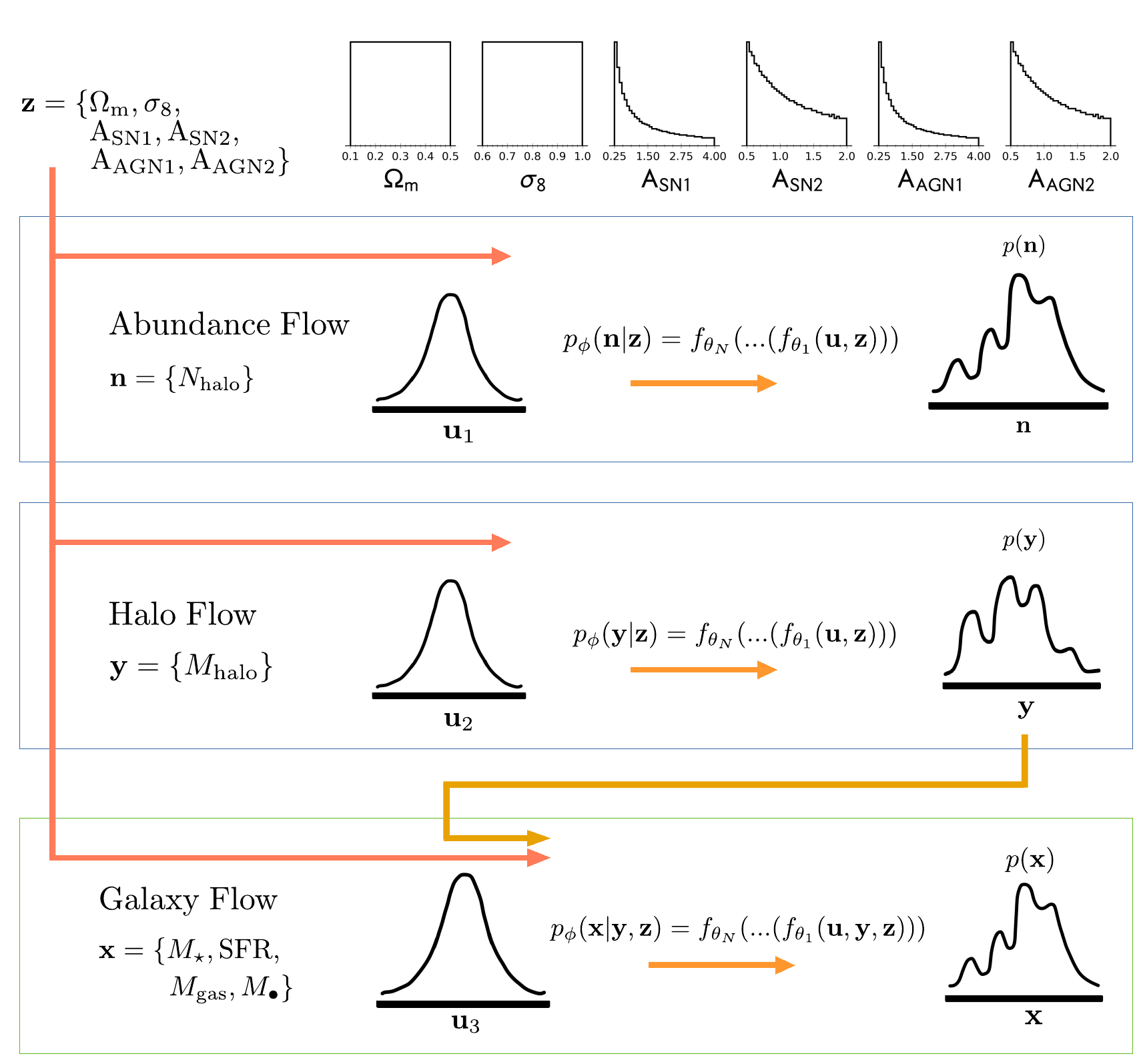}}
    \caption{High level diagram of the model. The distribution of the conditional cosmological and astrophysical parameters is shown at the top.
    The \textit{abundance}, \textit{halo} and \textit{galaxy flows} are shown below.
    The arrows highlight the direction of conditional dependence, as well as the mapping from each simple base distribution to the complex target density distribution.
    }
    \label{toy_diagram}
    \end{center}
    \vskip -0.2in
\end{figure*}

A normalizing flow (NF) models some data $\textbf{x}$ as a bijective transformation of some base distribution, typically a gaussian noise variable $\textbf{u}$,
\begin{align}
    \textbf{x} &= f_{\theta}(\textbf{u}) \\
    \textbf{u} &\sim \pi(\textbf{u}) \;,
\end{align}
where $f_{\theta}$ is invertible and differentiable, with parameters $\theta$.
This allows the target density $p_{\phi}(\textbf{x})$ to be written as
\begin{align}
    p_{\phi}(\textbf{x}) = \pi(f^{-1}_{\theta}(\textbf{x}))\left | \, \mathrm{det}\left (\frac{\delta f^{-1}_{\theta}}{\delta \textbf{x}} \right ) \,\right | \;\; .
\end{align}
For maximum flexibility $f_{\theta}$ and $f^{-1}_{\theta}$ are modelled using invertible neural networks (NN).
$f_{\theta}$ can be represented by multiple stacked layers, in order to produce highly complex mappings from the noise to the target density.

In order to build a conditional model we require a dataset with pairs of variables, $\mathcal{D} = \left\{ \left( \textbf{z}, \textbf{x} \right) \right\}$.
Here, the $\textbf{z}$ parameters are responsible for the generation of $\textbf{x}$, and we wish to model $p_{x}( \textbf{x} | \textbf{z} )$.
To include this conditional dependence in our model we incorporate these parameters in our transformation, $\textbf{x} = f_{\theta}(\textbf{u}, \textbf{z})$ \citep{winkler_learning_2019}.
We implement a version of a Neural spline flow \citep{durkan_neural_2019,dolatabadi_invertible_2020}.

The Cosmology and Astrophysics with MachinE Learning Simulations \citep[CAMELS;][]{villaescusa-navarro_camels_2021} are a large ensemble of $N$-body and hydrodynamic simulations exploring the effect of cosmological and astrophysical parameter choices on galaxy evolution and structure formation.
In this study we focus on the \textsc{Simba} simulation suite only \citep{dave_simba:_2019}.
For full details please refer to \citet{villaescusa-navarro_camels_2021,villaescusa-navarro_camels_2023,ni_camels_2023}.
Each simulation is defined by the initial random phases, as well as 4 astrophysical parameters ($A_{\mathrm{SN1}}$, $A_{\mathrm{SN2}}$, $A_{\mathrm{AGN1}}$, $A_{\mathrm{AGN2}}$) and 2 cosmological parameters ($\Omega_{\mathrm{m}}$, $\sigma_{8}$).
The following cosmological parameters are kept fixed in all simulations: $\Omega_{\mathrm{b}} = 0.049$, $h = 0.6711$, $n_s =
0.9624$, $M_{\nu} = 0.0 \; \mathrm{eV}$, $w = -1$, $\Omega_K = 0$.
The fiducial astrophysical parameters are defined at $A = 1.0$ and varied around this value to control the relative strength of the various feedback implementations in each simulation.
There are a number of different simulation sets within the CAMELS suite; the Latin Hypercube (LH) set contains 1000 simulations where the 6 parameters are varied using a latin hypercube; the cosmic variance CV set contains 27 simulations that only differ in the value of the random seed in the initial conditions.

We train three complementary flows, each conditional on the cosmological and astrophysical parameters  (an illustration of the different flows is shown in Figure~\ref{toy_diagram}).
The \textit{abundance flow} models the absolute abundance of subhaloes with mass $> 10^{10} \mathrm{M_{\odot}}$, $p_{\phi}(\textbf{n} \,|\, \textbf{z})$. 
We add gaussian noise to the data in the LH set equal to the scatter in the abundance in the CV set 50 times, and train on this augmented data set, to mimic the effect of cosmic variance.
The \textit{halo flow} models the density distribution of halo masses, $p_{\phi}(\textbf{y} \,|\, \textbf{z})$.
By coupling the \textit{abundance} and \textit{halo flows}, we can generate the volume normalised halo mass function for arbitrary parameters; an example is shown in the top left corner of Figure \ref{hmf_gsmf}.
Finally, the \textit{galaxy flow} models the distribution of galaxy properties within dark matter haloes by further conditioning on the subhalo mass, $p_{\phi}(\textbf{x} \,|\, \textbf{z, y})$.
We predict the stellar mass, gas mass, black hole mass and star formation rate.
%An illustration of the different flows is shown in Figure \ref{toy_diagram}.

We reserve a random subset of entire LH set simulations for testing (15\%), and use the rest for training and validation; this ensures there is no overlap between the train and test sets of galaxies with the same astrophysical and cosmological parameters.
We use the $z = 0$ snapshot from each simulation, and reserve a study of the redshift dependence for future work.
Each flow contains 16 layers, each consisting of a linear rational spline bijection (with 256 segments) coupled to an autoregressive NN layer consisting of two hidden layers with 256 and 128 nodes, respectively.
We train using the ADAM optimizer \citep{kingma_adam:_2014}, with a multi-step learning rate starting at $5 \times 10^{-3}$, with $\gamma = 0.1$, using mini batches of size 2048 that are randomly shuffled after each epoch.
At the end of each epoch we evaluate on the validation set, and save the model if the validation error has improved, to avoid overfitting.

\section{Results}

\begin{figure*}%[ht]
    \vskip 0.2in
    \begin{center}
    \centerline{\includegraphics[width=0.85\textwidth]{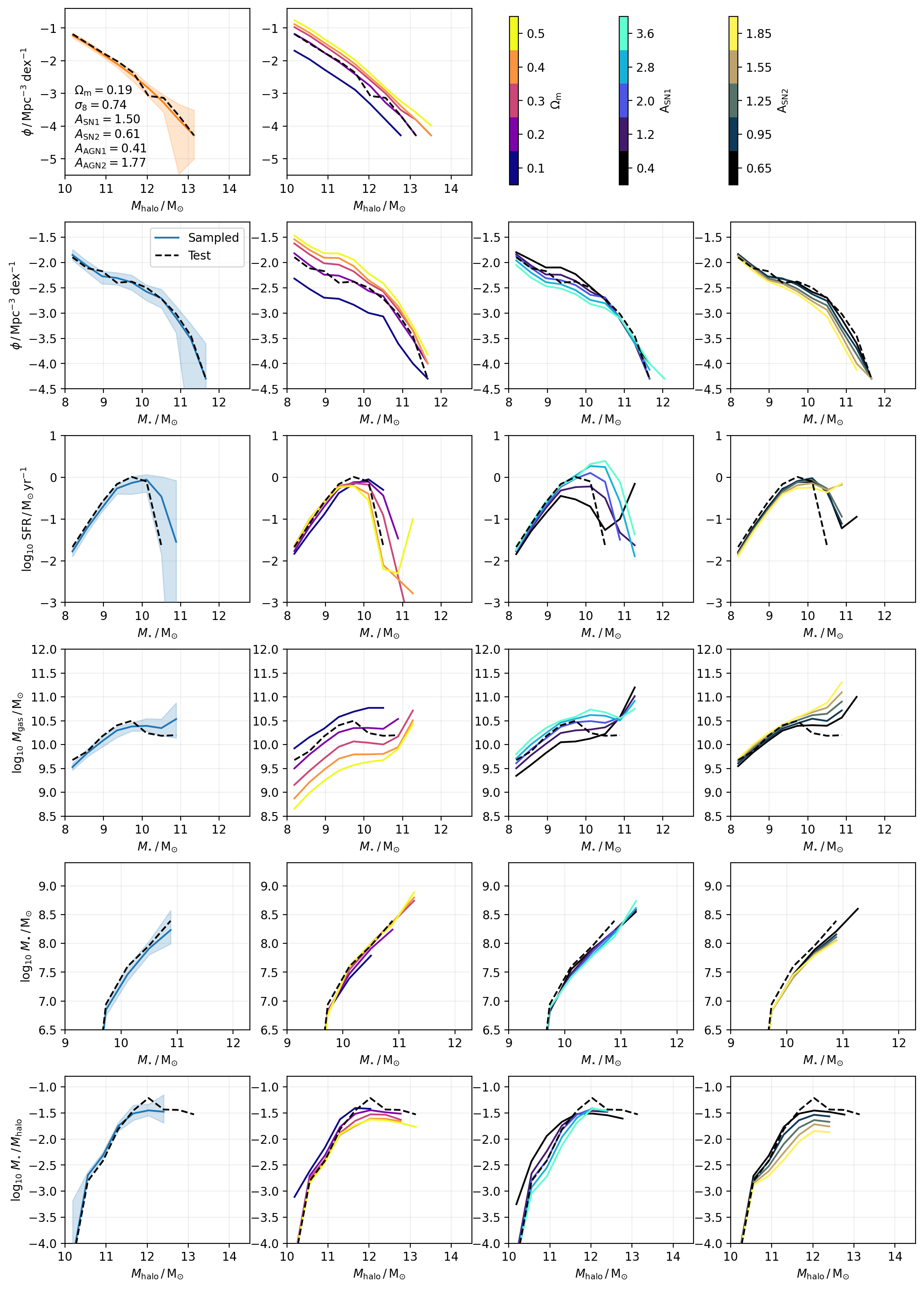}}
    \caption{An example of the model predictions when used as a generative model for haloes and galaxies, for fixed and varying parameters. 
    The top row shows the halo mass function (HMF), the second row the galaxy stellar mass function, the third row the star forming sequence, the fourth row the stellar mass--gas mass relation, the fifth row the stellar mass--black hole mass relation, and finally the stellar--halo mass relation.
    The first row shows predictions using haloes generated from the \textit{abundance} and \textit{halo flows}, as well as haloes taken directly from the LH set simulation.
    }
    \label{hmf_gsmf}
    \end{center}
    \vskip -0.2in
\end{figure*}

In this section we demonstrate an example use case for the model by predicting the galaxy and halo properties for a set of parameters not used in the training procedure.
We take these parameters from an LH set simulation from the test set, and first predict the halo mass function given the input parameters $\textbf{z}$.
We then use the \textit{abundance flow} to predict the cumulative number of subhaloes with mass $M_{\mathrm{halo}} > 10^{10} \, M_{\odot}$, $\textbf{n}$, and the \textit{halo flow} to predict the distribution of their masses, $\textbf{y}$.
Combined we can produce the halo mass function (HMF), shown in the top left panel of Figure \ref{hmf_gsmf} for 50 realisations, and compared to the true HMF from the corresponding LH set simulation.
The model successfully reproduces the distribution function within the scatter of the realisations.
We can also change one of the conditional parameters and explore the impact on the HMF.
This is shown in the top row of Figure \ref{hmf_gsmf}; there is a strong positive correlation between $\Omega_{\mathrm{m}}$ and the normalisation of the HMF.

We can also predict the properties of the galaxy within each host subhalo by providing the subhalo mass as well as the other conditional parameters to the \textit{galaxy flow}.
Whilst galaxy properties may be dependent on additional parameters as well as mass, the flow is able to model the full distribution of those properties at a given mass, marginalising over these unknown additional dependencies.
The first panel in the second row of Figure \ref{hmf_gsmf} shows the galaxy stellar mass function (GSMF) produced when applied to haloes generated from the \textit{abundance} and \textit{halo flows}.
The GSMF is reproduced within the scatter of the 50 realisations.
We can, again, fix parameters and explore the impact on the GSMF; we show this for $\Omega_{\mathrm{m}}$, $\mathrm{A_{SN1}}$ \& $\mathrm{A_{SN2}}$ in the second row of Figure \ref{hmf_gsmf}.

The third, fourth, fifth and sixth rows in Figure \ref{hmf_gsmf} also show predictions for the star forming sequence, the stellar mass--gas mass relation, the stellar mass--black hole mass relation, and the stellar--halo mass relation, and the impact of changing conditional parameters ($\Omega_{\mathrm{m}}$, $A_{\mathrm{SN1}}$, $A_{\mathrm{SN2}}$) on each of these relations in turn.
We emphasise that galaxy properties are predicted jointly, enabling us to predict these relations self consistently.

\section{Conclusions}

We present a novel approach to modelling the galaxy--halo relationship, using the density estimation capabilities of normalising flows to model the coupled halo and galaxy distribution conditioned on astrophysical and cosmological parameters. 
The model is able to self-consistently predict a number of halo and galaxy relations, and shows interesting correlations with different cosmological and astrophysical parameters, whilst marginalising over other nuisance parameters.
There are a number of applications for such a model, from rapid generation of galaxy properties in dark matter only $N$-body simulations, to direct and indirect inference of astrophysical and cosmological parameters from individual galaxy properties or predicted scaling relations through simulation based inference (SBI) \citep{cranmer_modeling_2019}, an increasingly popular and flexible approach to inference \citep{papamakarios_masked_2017,alsing_fast_2019,hahn_likelihood_2019,zhang_real-time_2021,dax_real-time_2021,hahn_accelerated_2022,huppenkothen_accurate_2022,wang_sbi_2023}.

\section{Acknowledgements}
CCL acknowledges support from a Dennis Sciama fellowship funded by the University of Portsmouth for the Institute of Cosmology and Gravitation. DAA acknowledges support by NSF grants AST-2009687 and AST-2108944, CXO grant TM2-23006X, Simons Foundation Award CCA-1018464, and Cottrell Scholar Award CS-CSA-2023-028 by the Research Corporation for Science Advancement. GF acknowledges the support of the European Research Council under the Marie Sk\l{}odowska Curie actions through the Individual Global Fellowship No.~892401 PiCOGAMBAS and of the Simons Foundation.
NSMS acknowledges financial support from FAPESP, grants
\href{https://bv.fapesp.br/en/bolsas/187647/cosmological-covariance-matrices-and-machine-learning-methods/}{2019/13108-0} 
and \href{https://bv.fapesp.br/en/bolsas/202438/machine-learning-methods-for-extracting-cosmological-information/}{2022/03589-4}.

\bibliography{Normalizing_flows}

\begin{thebibliography}{39}
\providecommand{\natexlab}[1]{#1}
\providecommand{\url}[1]{\texttt{#1}}
\expandafter\ifx\csname urlstyle\endcsname\relax
  \providecommand{\doi}[1]{doi: #1}\else
  \providecommand{\doi}{doi: \begingroup \urlstyle{rm}\Url}\fi

\bibitem[Agarwal et~al.(2018)Agarwal, Davé, and Bassett]{agarwal_painting_2018}
Agarwal, S., Davé, R., and Bassett, B.~A.
\newblock Painting galaxies into dark matter haloes using machine learning.
\newblock \emph{MNRAS}, 478:\penalty0 3410--3422, August 2018.
\newblock ISSN 0035-8711.
\newblock \doi{10.1093/mnras/sty1169}.
\newblock URL \url{http://adsabs.harvard.edu/abs/2018MNRAS.478.3410A}.

\bibitem[Alsing et~al.(2019)Alsing, Charnock, Feeney, and Wandelt]{alsing_fast_2019}
Alsing, J., Charnock, T., Feeney, S., and Wandelt, B.
\newblock Fast likelihood-free cosmology with neural density estimators and active learning.
\newblock \emph{MNRAS}, 488:\penalty0 4440--4458, September 2019.
\newblock ISSN 0035-8711.
\newblock \doi{10.1093/mnras/stz1960}.
\newblock URL \url{https://ui.adsabs.harvard.edu/abs/2019MNRAS.488.4440A}.
\newblock ADS Bibcode: 2019MNRAS.488.4440A.

\bibitem[Benson(2010)]{benson_galaxy_2010}
Benson, A.~J.
\newblock Galaxy {Formation} {Theory}.
\newblock \emph{Phys. Rep.}, 495\penalty0 (2-3):\penalty0 33--86, October 2010.
\newblock ISSN 03701573.
\newblock \doi{10.1016/j.physrep.2010.06.001}.
\newblock URL \url{http://arxiv.org/abs/1006.5394}.
\newblock arXiv: 1006.5394.

\bibitem[Borrow et~al.(2022)Borrow, Schaller, Bahe, Schaye, Ludlow, Ploeckinger, Nobels, and Altamura]{borrow_impact_2022}
Borrow, J., Schaller, M., Bahe, Y.~M., Schaye, J., Ludlow, A.~D., Ploeckinger, S., Nobels, F. S.~J., and Altamura, E.
\newblock The impact of stochastic modeling on the predictive power of galaxy formation simulations.
\newblock \emph{arXiv.2211.08442}, November 2022.
\newblock \doi{10.48550/arXiv.2211.08442}.
\newblock URL \url{http://arxiv.org/abs/2211.08442}.
\newblock arXiv:2211.08442 [astro-ph].

\bibitem[Chittenden \& Tojeiro(2023)Chittenden and Tojeiro]{chittenden_modelling_2023}
Chittenden, H.~G. and Tojeiro, R.
\newblock Modelling the galaxy-halo connection with semi-recurrent neural networks.
\newblock \emph{MNRAS}, 518:\penalty0 5670--5692, January 2023.
\newblock ISSN 0035-8711.
\newblock \doi{10.1093/mnras/stac3498}.
\newblock URL \url{https://ui.adsabs.harvard.edu/abs/2023MNRAS.518.5670C}.
\newblock ADS Bibcode: 2023MNRAS.518.5670C.

\bibitem[Cranmer et~al.(2019)Cranmer, Galvez, Anderson, Spergel, and Ho]{cranmer_modeling_2019}
Cranmer, M.~D., Galvez, R., Anderson, L., Spergel, D.~N., and Ho, S.
\newblock Modeling the {Gaia} {Color}-{Magnitude} {Diagram} with {Bayesian} {Neural} {Flows} to {Constrain} {Distance} {Estimates}.
\newblock \emph{arXiv.1908.08045}, August 2019.
\newblock \doi{10.48550/arXiv.1908.08045}.
\newblock URL \url{https://ui.adsabs.harvard.edu/abs/2019arXiv190808045C}.
\newblock Publication Title: arXiv e-prints ADS Bibcode: 2019arXiv190808045C Type: article.

\bibitem[Davé et~al.(2019)Davé, Anglés-Alcázar, Narayanan, Li, Rafieferantsoa, and Appleby]{dave_simba:_2019}
Davé, R., Anglés-Alcázar, D., Narayanan, D., Li, Q., Rafieferantsoa, M.~H., and Appleby, S.
\newblock {SIMBA}: {Cosmological} simulations with black hole growth and feedback.
\newblock \emph{MNRAS}, 486\penalty0 (2):\penalty0 2827, June 2019.
\newblock \doi{10.1093/mnras/stz937}.
\newblock URL \url{https://ui.adsabs.harvard.edu/abs/2019MNRAS.486.2827D/abstract}.

\bibitem[Dax et~al.(2021)Dax, Green, Gair, Macke, Buonanno, and Schölkopf]{dax_real-time_2021}
Dax, M., Green, S.~R., Gair, J., Macke, J.~H., Buonanno, A., and Schölkopf, B.
\newblock Real-{Time} {Gravitational} {Wave} {Science} with {Neural} {Posterior} {Estimation}.
\newblock \emph{Phys. Rev. L}, 127:\penalty0 241103, December 2021.
\newblock ISSN 0031-9007.
\newblock \doi{10.1103/PhysRevLett.127.241103}.
\newblock URL \url{https://ui.adsabs.harvard.edu/abs/2021PhRvL.127x1103D}.
\newblock ADS Bibcode: 2021PhRvL.127x1103D.

\bibitem[de~Santi et~al.(2022)de~Santi, Rodrigues, Montero-Dorta, Abramo, Tucci, and Artale]{de_santi_mimicking_2022}
de~Santi, N. S.~M., Rodrigues, N. V.~N., Montero-Dorta, A.~D., Abramo, L.~R., Tucci, B., and Artale, M.~C.
\newblock Mimicking the halo-galaxy connection using machine learning.
\newblock \emph{MNRAS}, 514:\penalty0 2463--2478, August 2022.
\newblock ISSN 0035-8711.
\newblock \doi{10.1093/mnras/stac1469}.
\newblock URL \url{https://ui.adsabs.harvard.edu/abs/2022MNRAS.514.2463D}.
\newblock ADS Bibcode: 2022MNRAS.514.2463D.

\bibitem[Dinh et~al.(2015)Dinh, Krueger, and Bengio]{dinh_nice_2015}
Dinh, L., Krueger, D., and Bengio, Y.
\newblock {NICE}: {Non}-linear {Independent} {Components} {Estimation}.
\newblock In \emph{{ICLR}}, 2015.
\newblock URL \url{https://ui.adsabs.harvard.edu/abs/2014arXiv1410.8516D}.

\bibitem[Dolatabadi et~al.(2020)Dolatabadi, Erfani, and Leckie]{dolatabadi_invertible_2020}
Dolatabadi, H.~M., Erfani, S., and Leckie, C.
\newblock Invertible {Generative} {Modeling} using {Linear} {Rational} {Splines}.
\newblock In \emph{23rd {International} {Conference} on {Artificial} {Intelligence} and {Statistics} ({AISTATS})}, January 2020.
\newblock URL \url{https://ui.adsabs.harvard.edu/abs/2020arXiv200105168D}.

\bibitem[Durkan et~al.(2019)Durkan, Bekasov, Murray, and Papamakarios]{durkan_neural_2019}
Durkan, C., Bekasov, A., Murray, I., and Papamakarios, G.
\newblock Neural {Spline} {Flows}.
\newblock In \emph{33rd {Conference} on {Neural} {Information} {Processing} {Systems} ({NeurIPS} 2019)}, June 2019.
\newblock URL \url{https://ui.adsabs.harvard.edu/abs/2019arXiv190604032D}.

\bibitem[Friedman \& Hassan(2022)Friedman and Hassan]{friedman_higlow_2022}
Friedman, R. and Hassan, S.
\newblock {HIGlow}: {Conditional} {Normalizing} {Flows} for {High}-{Fidelity} {HI} {Map} {Modeling}.
\newblock In \emph{Machine {Learning} and the {Physical} {Sciences} workshop, {NeurIPS} 2022}, November 2022.
\newblock URL \url{https://ui.adsabs.harvard.edu/abs/2022arXiv221112724F}.

\bibitem[Genel et~al.(2019)Genel, Bryan, Springel, Hernquist, Nelson, Pillepich, Weinberger, Pakmor, Marinacci, and Vogelsberger]{genel_quantification_2019}
Genel, S., Bryan, G.~L., Springel, V., Hernquist, L., Nelson, D., Pillepich, A., Weinberger, R., Pakmor, R., Marinacci, F., and Vogelsberger, M.
\newblock A {Quantification} of the {Butterfly} {Effect} in {Cosmological} {Simulations} and {Implications} for {Galaxy} {Scaling} {Relations}.
\newblock \emph{The Astrophysical Journal}, 871:\penalty0 21, January 2019.
\newblock ISSN 0004-637X.
\newblock \doi{10.3847/1538-4357/aaf4bb}.
\newblock URL \url{https://ui.adsabs.harvard.edu/abs/2019ApJ...871...21G}.
\newblock ADS Bibcode: 2019ApJ...871...21G.

\bibitem[Goodfellow et~al.(2014)Goodfellow, Pouget-Abadie, Mirza, Xu, Warde-Farley, Ozair, Courville, and Bengio]{goodfellow_generative_2014}
Goodfellow, I., Pouget-Abadie, J., Mirza, M., Xu, B., Warde-Farley, D., Ozair, S., Courville, A., and Bengio, Y.
\newblock Generative {Adversarial} {Nets}.
\newblock In \emph{Advances in {Neural} {Information} {Processing} {Systems}}, volume~27, 2014.
\newblock URL \url{https://papers.nips.cc/paper_files/paper/2014/hash/5ca3e9b122f61f8f06494c97b1afccf3-Abstract.html}.

\bibitem[Hahn \& Melchior(2022)Hahn and Melchior]{hahn_accelerated_2022}
Hahn, C. and Melchior, P.
\newblock Accelerated {Bayesian} {SED} {Modeling} {Using} {Amortized} {Neural} {Posterior} {Estimation}.
\newblock \emph{ApJ}, 938:\penalty0 11, October 2022.
\newblock ISSN 0004-637X.
\newblock \doi{10.3847/1538-4357/ac7b84}.
\newblock URL \url{https://ui.adsabs.harvard.edu/abs/2022ApJ...938...11H}.
\newblock ADS Bibcode: 2022ApJ...938...11H.

\bibitem[Hahn et~al.(2019)Hahn, Beutler, Sinha, Berlind, Ho, and Hogg]{hahn_likelihood_2019}
Hahn, C., Beutler, F., Sinha, M., Berlind, A., Ho, S., and Hogg, D.~W.
\newblock Likelihood non-{Gaussianity} in large-scale structure analyses.
\newblock \emph{MNRAS}, 485:\penalty0 2956--2969, May 2019.
\newblock ISSN 0035-8711.
\newblock \doi{10.1093/mnras/stz558}.
\newblock URL \url{https://ui.adsabs.harvard.edu/abs/2019MNRAS.485.2956H}.
\newblock ADS Bibcode: 2019MNRAS.485.2956H.

\bibitem[Hassan et~al.(2022)Hassan, Villaescusa-Navarro, Wandelt, Spergel, Anglés-Alcázar, Genel, Cranmer, Bryan, Davé, Somerville, Eickenberg, Narayanan, Ho, and Andrianomena]{hassan_hiflow_2022}
Hassan, S., Villaescusa-Navarro, F., Wandelt, B., Spergel, D.~N., Anglés-Alcázar, D., Genel, S., Cranmer, M., Bryan, G.~L., Davé, R., Somerville, R.~S., Eickenberg, M., Narayanan, D., Ho, S., and Andrianomena, S.
\newblock {HIFLOW}: {Generating} {Diverse} {HI} {Maps} and {Inferring} {Cosmology} while {Marginalizing} over {Astrophysics} {Using} {Normalizing} {Flows}.
\newblock \emph{ApJ}, 937:\penalty0 83, October 2022.
\newblock ISSN 0004-637X.
\newblock \doi{10.3847/1538-4357/ac8b09}.
\newblock URL \url{https://ui.adsabs.harvard.edu/abs/2022ApJ...937...83H}.
\newblock ADS Bibcode: 2022ApJ...937...83H.

\bibitem[Huppenkothen \& Bachetti(2022)Huppenkothen and Bachetti]{huppenkothen_accurate_2022}
Huppenkothen, D. and Bachetti, M.
\newblock Accurate {X}-ray timing in the presence of systematic biases with simulation-based inference.
\newblock \emph{MNRAS}, 511:\penalty0 5689--5708, April 2022.
\newblock ISSN 0035-8711.
\newblock \doi{10.1093/mnras/stab3437}.
\newblock URL \url{https://ui.adsabs.harvard.edu/abs/2022MNRAS.511.5689H}.
\newblock ADS Bibcode: 2022MNRAS.511.5689H.

\bibitem[Icaza-Lizaola et~al.(2023)Icaza-Lizaola, Bower, Norberg, Cole, and Schaller]{icaza-lizaola_sparse_2023}
Icaza-Lizaola, M., Bower, R.~G., Norberg, P., Cole, S., and Schaller, M.
\newblock A sparse regression approach for populating dark matter haloes and subhaloes with galaxies.
\newblock \emph{MNRAS}, 518:\penalty0 2903--2920, January 2023.
\newblock ISSN 0035-8711.
\newblock \doi{10.1093/mnras/stac3265}.
\newblock URL \url{https://ui.adsabs.harvard.edu/abs/2023MNRAS.518.2903I}.
\newblock ADS Bibcode: 2023MNRAS.518.2903I.

\bibitem[Jespersen et~al.(2022)Jespersen, Cranmer, Melchior, Ho, Somerville, and Gabrielpillai]{jespersen_mangrove_2022}
Jespersen, C.~K., Cranmer, M., Melchior, P., Ho, S., Somerville, R.~S., and Gabrielpillai, A.
\newblock Mangrove: {Learning} {Galaxy} {Properties} from {Merger} {Trees}.
\newblock \emph{ApJ}, 941:\penalty0 7, December 2022.
\newblock ISSN 0004-637X.
\newblock \doi{10.3847/1538-4357/ac9b18}.
\newblock URL \url{https://ui.adsabs.harvard.edu/abs/2022ApJ...941....7J}.
\newblock ADS Bibcode: 2022ApJ...941....7J.

\bibitem[Jimenez~Rezende \& Mohamed(2015)Jimenez~Rezende and Mohamed]{jimenez_rezende_variational_2015}
Jimenez~Rezende, D. and Mohamed, S.
\newblock Variational {Inference} with {Normalizing} {Flows}.
\newblock In \emph{Proceedings of the 32nd {International} {Conference} on {Machine} {Learning}}, May 2015.
\newblock URL \url{https://ui.adsabs.harvard.edu/abs/2015arXiv150505770J}.

\bibitem[Jimenez~Rezende et~al.(2014)Jimenez~Rezende, Mohamed, and Wierstra]{jimenez_rezende_stochastic_2014}
Jimenez~Rezende, D., Mohamed, S., and Wierstra, D.
\newblock Stochastic {Backpropagation} and {Approximate} {Inference} in {Deep} {Generative} {Models}.
\newblock In \emph{Proceedings of the 31st {International} {Conference} on {Machine} {Learning} ({ICML})}, January 2014.
\newblock URL \url{https://ui.adsabs.harvard.edu/abs/2014arXiv1401.4082J}.

\bibitem[Jo \& Kim(2019)Jo and Kim]{jo_machine-assisted_2019}
Jo, Y. and Kim, J.-h.
\newblock Machine-assisted semi-simulation model ({MSSM}): estimating galactic baryonic properties from their dark matter using a machine trained on hydrodynamic simulations.
\newblock \emph{MNRAS}, 489:\penalty0 3565--3581, November 2019.
\newblock ISSN 0035-8711.
\newblock \doi{10.1093/mnras/stz2304}.
\newblock URL \url{http://adsabs.harvard.edu/abs/2019MNRAS.489.3565J}.

\bibitem[Kamdar et~al.(2016)Kamdar, Turk, and Brunner]{kamdar_machine_2016}
Kamdar, H.~M., Turk, M.~J., and Brunner, R.~J.
\newblock Machine learning and cosmological simulations – {I}. {Semi}-analytical models.
\newblock \emph{MNRAS}, 455\penalty0 (1):\penalty0 642--658, January 2016.
\newblock ISSN 0035-8711, 1365-2966.
\newblock \doi{10.1093/mnras/stv2310}.
\newblock URL \url{http://mnras.oxfordjournals.org/content/455/1/642}.

\bibitem[Kingma \& Ba(2014)Kingma and Ba]{kingma_adam:_2014}
Kingma, D.~P. and Ba, J.
\newblock Adam: {A} {Method} for {Stochastic} {Optimization}.
\newblock In \emph{3rd {International} {Conference} for {Learning} {Representations}}, December 2014.
\newblock URL \url{http://adsabs.harvard.edu/abs/2014arXiv1412.6980K}.

\bibitem[Kingma \& Dhariwal(2018)Kingma and Dhariwal]{kingma_glow_2018}
Kingma, D.~P. and Dhariwal, P.
\newblock Glow: {Generative} {Flow} with {Invertible} 1x1 {Convolutions}.
\newblock In \emph{Advances in {Neural} {Information} {Processing} {Systems}}, July 2018.
\newblock URL \url{http://arxiv.org/abs/1807.03039}.
\newblock arXiv:1807.03039 [cs, stat].

\bibitem[Kingma \& Welling(2013)Kingma and Welling]{kingma_auto-encoding_2013}
Kingma, D.~P. and Welling, M.
\newblock Auto-{Encoding} {Variational} {Bayes}.
\newblock In \emph{2nd {International} {Conference} on {Learning} {Representations} ({ICLR2014})}, December 2013.
\newblock URL \url{https://ui.adsabs.harvard.edu/abs/2013arXiv1312.6114K}.

\bibitem[Lovell et~al.(2022)Lovell, Wilkins, Thomas, Schaller, Baugh, Fabbian, and Bahé]{lovell_machine_2022}
Lovell, C.~C., Wilkins, S.~M., Thomas, P.~A., Schaller, M., Baugh, C.~M., Fabbian, G., and Bahé, Y.
\newblock A machine learning approach to mapping baryons on to dark matter haloes using the {EAGLE} and {C}-{EAGLE} simulations.
\newblock \emph{MNRAS}, 509:\penalty0 5046--5061, February 2022.
\newblock ISSN 0035-8711.
\newblock \doi{10.1093/mnras/stab3221}.
\newblock URL \url{https://ui.adsabs.harvard.edu/abs/2022MNRAS.509.5046L}.
\newblock ADS Bibcode: 2022MNRAS.509.5046L.

\bibitem[Ni et~al.(2023)Ni, Genel, Anglés-Alcázar, Villaescusa-Navarro, Jo, Bird, Di~Matteo, Croft, Chen, de~Santi, Gebhardt, Shao, Pandey, Hernquist, and Dave]{ni_camels_2023}
Ni, Y., Genel, S., Anglés-Alcázar, D., Villaescusa-Navarro, F., Jo, Y., Bird, S., Di~Matteo, T., Croft, R., Chen, N., de~Santi, N. S.~M., Gebhardt, M., Shao, H., Pandey, S., Hernquist, L., and Dave, R.
\newblock The {CAMELS} project: {Expanding} the galaxy formation model space with new {ASTRID} and 28-parameter {TNG} and {SIMBA} suites.
\newblock \emph{arXiv.2304.02096}, April 2023.
\newblock \doi{10.48550/arXiv.2304.02096}.
\newblock URL \url{https://ui.adsabs.harvard.edu/abs/2023arXiv230402096N}.
\newblock Publication Title: arXiv e-prints ADS Bibcode: 2023arXiv230402096N Type: article.

\bibitem[Papamakarios et~al.(2017)Papamakarios, Pavlakou, and Murray]{papamakarios_masked_2017}
Papamakarios, G., Pavlakou, T., and Murray, I.
\newblock Masked {Autoregressive} {Flow} for {Density} {Estimation}.
\newblock In Guyon, I., Luxburg, U.~V., Bengio, S., Wallach, H., Fergus, R., Vishwanathan, S., and Garnett, R. (eds.), \emph{Advances in {Neural} {Information} {Processing} {Systems}}, 2017.
\newblock URL \url{https://proceedings.neurips.cc/paper_files/paper/2017/file/6c1da886822c67822bcf3679d04369fa-Paper.pdf}.

\bibitem[Rodrigues et~al.(2023)Rodrigues, de~Santi, Montero-Dorta, and Abramo]{rodrigues_high-fidelity_2023}
Rodrigues, N. V.~N., de~Santi, N. S.~M., Montero-Dorta, A.~D., and Abramo, L.~R.
\newblock High-fidelity reproduction of central galaxy joint distributions with {Neural} {Networks}.
\newblock Technical report, January 2023.
\newblock URL \url{https://ui.adsabs.harvard.edu/abs/2023arXiv230106398R}.
\newblock Publication Title: arXiv e-prints ADS Bibcode: 2023arXiv230106398R Type: article.

\bibitem[Somerville \& Davé(2015)Somerville and Davé]{somerville_physical_2015}
Somerville, R.~S. and Davé, R.
\newblock Physical {Models} of {Galaxy} {Formation} in a {Cosmological} {Framework}.
\newblock \emph{ARAA}, 53\penalty0 (1):\penalty0 51--113, August 2015.
\newblock ISSN 0066-4146, 1545-4282.
\newblock \doi{10.1146/annurev-astro-082812-140951}.
\newblock URL \url{http://arxiv.org/abs/1412.2712}.
\newblock arXiv: 1412.2712.

\bibitem[Villaescusa-Navarro et~al.(2021)Villaescusa-Navarro, Anglés-Alcázar, Genel, Spergel, Somerville, Dave, Pillepich, Hernquist, Nelson, Torrey, Narayanan, Li, Philcox, La~Torre, Maria~Delgado, Ho, Hassan, Burkhart, Wadekar, Battaglia, Contardo, and Bryan]{villaescusa-navarro_camels_2021}
Villaescusa-Navarro, F., Anglés-Alcázar, D., Genel, S., Spergel, D.~N., Somerville, R.~S., Dave, R., Pillepich, A., Hernquist, L., Nelson, D., Torrey, P., Narayanan, D., Li, Y., Philcox, O., La~Torre, V., Maria~Delgado, A., Ho, S., Hassan, S., Burkhart, B., Wadekar, D., Battaglia, N., Contardo, G., and Bryan, G.~L.
\newblock The {CAMELS} {Project}: {Cosmology} and {Astrophysics} with {Machine}-learning {Simulations}.
\newblock \emph{ApJ}, 915:\penalty0 71, July 2021.
\newblock ISSN 0004-637X.
\newblock \doi{10.3847/1538-4357/abf7ba}.
\newblock URL \url{https://ui.adsabs.harvard.edu/abs/2021ApJ...915...71V}.
\newblock ADS Bibcode: 2021ApJ...915...71V.

\bibitem[Villaescusa-Navarro et~al.(2023)Villaescusa-Navarro, Genel, Anglés-Alcázar, Perez, Villanueva-Domingo, Wadekar, Shao, Mohammad, Hassan, Moser, Lau, Machado Poletti~Valle, Nicola, Thiele, Jo, Philcox, Oppenheimer, Tillman, Hahn, Kaushal, Pisani, Gebhardt, Delgado, Caliendo, Kreisch, Wong, Coulton, Eickenberg, Parimbelli, Ni, Steinwandel, La~Torre, Dave, Battaglia, Nagai, Spergel, Hernquist, Burkhart, Narayanan, Wandelt, Somerville, Bryan, Viel, Li, Irsic, Kraljic, Marinacci, and Vogelsberger]{villaescusa-navarro_camels_2023}
Villaescusa-Navarro, F., Genel, S., Anglés-Alcázar, D., Perez, L.~A., Villanueva-Domingo, P., Wadekar, D., Shao, H., Mohammad, F.~G., Hassan, S., Moser, E., Lau, E.~T., Machado Poletti~Valle, L.~F., Nicola, A., Thiele, L., Jo, Y., Philcox, O. H.~E., Oppenheimer, B.~D., Tillman, M., Hahn, C., Kaushal, N., Pisani, A., Gebhardt, M., Delgado, A.~M., Caliendo, J., Kreisch, C., Wong, K. W.~K., Coulton, W.~R., Eickenberg, M., Parimbelli, G., Ni, Y., Steinwandel, U.~P., La~Torre, V., Dave, R., Battaglia, N., Nagai, D., Spergel, D.~N., Hernquist, L., Burkhart, B., Narayanan, D., Wandelt, B., Somerville, R.~S., Bryan, G.~L., Viel, M., Li, Y., Irsic, V., Kraljic, K., Marinacci, F., and Vogelsberger, M.
\newblock The {CAMELS} {Project}: {Public} {Data} {Release}.
\newblock \emph{ApJSS}, 265:\penalty0 54, April 2023.
\newblock ISSN 0067-0049.
\newblock \doi{10.3847/1538-4365/acbf47}.
\newblock URL \url{https://ui.adsabs.harvard.edu/abs/2023ApJS..265...54V}.
\newblock ADS Bibcode: 2023ApJS..265...54V.

\bibitem[Wang et~al.(2023)Wang, Leja, Villar, and Speagle]{wang_sbi_2023}
Wang, B., Leja, J., Villar, V.~A., and Speagle, J.~S.
\newblock {SBI}++: {Flexible}, {Ultra}-fast {Likelihood}-free {Inference} {Customized} for {Astronomical} {Application}.
\newblock \emph{arXiv.2304.05281}, April 2023.
\newblock \doi{10.48550/arXiv.2304.05281}.
\newblock URL \url{https://ui.adsabs.harvard.edu/abs/2023arXiv230405281W}.
\newblock Publication Title: arXiv e-prints ADS Bibcode: 2023arXiv230405281W Type: article.

\bibitem[Wechsler \& Tinker(2018)Wechsler and Tinker]{wechsler_connection_2018}
Wechsler, R.~H. and Tinker, J.~L.
\newblock The {Connection} {Between} {Galaxies} and {Their} {Dark} {Matter} {Halos}.
\newblock \emph{ARAA}, 56\penalty0 (1):\penalty0 435--487, 2018.
\newblock \doi{10.1146/annurev-astro-081817-051756}.
\newblock URL \url{https://doi.org/10.1146/annurev-astro-081817-051756}.
\newblock \_eprint: https://doi.org/10.1146/annurev-astro-081817-051756.

\bibitem[Winkler et~al.(2019)Winkler, Worrall, Hoogeboom, and Welling]{winkler_learning_2019}
Winkler, C., Worrall, D., Hoogeboom, E., and Welling, M.
\newblock Learning {Likelihoods} with {Conditional} {Normalizing} {Flows}.
\newblock \emph{arXiv.1912.00042}, November 2019.
\newblock \doi{10.48550/arXiv.1912.00042}.
\newblock URL \url{https://ui.adsabs.harvard.edu/abs/2019arXiv191200042W}.
\newblock Publication Title: arXiv e-prints ADS Bibcode: 2019arXiv191200042W Type: article.

\bibitem[Zhang et~al.(2021)Zhang, Bloom, Gaudi, Lanusse, Lam, and Lu]{zhang_real-time_2021}
Zhang, K., Bloom, J.~S., Gaudi, B.~S., Lanusse, F., Lam, C., and Lu, J.~R.
\newblock Real-time {Likelihood}-free {Inference} of {Roman} {Binary} {Microlensing} {Events} with {Amortized} {Neural} {Posterior} {Estimation}.
\newblock \emph{AJ}, 161\penalty0 (6):\penalty0 262, May 2021.
\newblock ISSN 1538-3881.
\newblock \doi{10.3847/1538-3881/abf42e}.
\newblock URL \url{https://dx.doi.org/10.3847/1538-3881/abf42e}.
\newblock Publisher: The American Astronomical Society.

\end{thebibliography}
\bibliographystyle{icml2023}

\end{document}